\documentclass[journal]{IEEEtran}

\usepackage[dvipsnames]{xcolor}
\usepackage{graphicx}
\usepackage[caption=false]{subfig}

\ifCLASSINFOpdf
\else
\fi
\usepackage{amsmath,amssymb}
\usepackage{url}

\begin{document}
%
\title{One-dimensional $p$-wave superconductor toy-model for Majorana fermions in multiband semiconductor nanowires}
%
%
%

\author{Ant\^{o}nio~L.~R.~Manesco,
        Gabriel~Weber,
        and~Durval~Rodrigues~Jr.
\thanks{Ant\^{o}nio~L.~R.~Manesco, Gabriel~Weber, and~Durval~Rodrigues~Jr. are with the Lorena Engineering School, University of S\~{a}o Paulo, Lorena, SP, 12600-970 Brazil (e-mail: durval@demar.eel.usp.br, antoniolrm@usp.br).}
\thanks{The work of ALRM was supported by FAPESP grant No. 2016/10167-8.}
\thanks{DRJ is a CNPq researcher.}
\thanks{Manuscript received September 19, 2017.}
\thanks{Accepted version. Citation information: DOI 10.1109/TASC.2018.2807361, IEEE Transactions on Applied Superconductivity.}
}

%
%

\markboth{}%
{Shell \MakeLowercase{\textit{et al.}}: Bare Demo of IEEEtran.cls for IEEE Journals}
%



\maketitle

\begin{abstract}
Majorana fermions are particles identical to their antiparticles proposed theoretically in 1937 by Ettore Majorana as real solutions of the Dirac equation. Alexei Kitaev suggested that Majorana particles should emerge in condensed matter systems as zero mode excitations in one-dimensional $p$-wave superconductors, with possible applications in quantum computation due to their non-abelian statistics. The search for Majorana zero modes in condensed matter systems led to one of the first realistic models based in a semiconductor nanowire with high spin-orbit coupling, induced superconducting $s$-wave pairing and Zeeman splitting. Soon, it was realized that size-quantization effects should generate subbands in these systems that could even allow the emergence of more than one Majorana mode at each edge, resulting in a zero bias peak on the differential conductance with a different shape from the predicted by simplified theoretical models. In this work, we provide a connection between a finite-size nanowire with two occupied subbands and a 2-band Kitaev chain, and discuss the advantage of a one-dimensional model to understand the phenomenology of the system, including the presence of a hidden chiral symmetry and its similarity with a spinfull Kitaev chain under a magnetic field.
\end{abstract}

Topological superconductivity, Majorana modes, semiconductor nanowire, quantum device.

%
\IEEEpeerreviewmaketitle

\section{Introduction}
%
%
%
%
\IEEEPARstart{Q}{uantum} computation presents a theoretical framework for efficiently solving hard computational tasks, \textit{e.g.}, factorization of large numbers using Shor's algorithm \cite{shor1999polynomial}. However, a practical problem in a system with many quantum bits (qubits) is quantum decoherence. The interaction with an external environment collapses the wavefunction and no reliable information can be obtained. In that sense, an error-free quantum computer needs to somehow avoid decoherence phenomena.

While in classical computers it is necessary to implement error correction, an alternative route on the quantum case for a fault-tolerant computer is the use of topological quantum computation \cite{sarma2006topological,pachos2012introduction}. The main idea is to construct systems in which the presence of discrete symmetries protect the prepared states, making the system robust to decoherence.

A final ingredient needed in a topological quantum computer is a way of introducing unitary transformations as logic quantum gates. This can be done with non-abelian braiding statistics \cite{sarma2006topological,pachos2012introduction}. Thus, topologically protected condensed matter systems with non-abelian anyonic quasi-particles provides a suitable scenario for practical implementation of topological quantum computation.

As introduced in a seminal paper by Alexei Kitaev, a one-dimensional spinless $p$-wave superconductor (a.k.a. Kitaev chain) can host end modes that are non-abelian anyonic Majorana quasi-particles \cite{kitaev2001unpaired}. The name ``Majorana'' refers to a special class of particles, first considered by Ettore Majorana in the context of high-energy physics, that are their own antiparticles as a result of purely imaginary gamma-matrices on the Dirac equation \cite{majorana1937teoria}. To present such behavior, the particles need to have no electrical charge. In superconductors, the Bogoliubov quasi-particle excitations consist of linear combinations of electron and holes, hence, zero energy excitations have no charge. Since the Kitaev chain can present non-trivial topology, it is expected the emergence of zero-energy end states, which are Majorana non-abelian anyons \cite{kitaev2003fault, sato2016topological}.

Due to the interest both in purely scientific as well as technological applications, a large number of realistic systems realizing the phenomenology of the Kitaev chain were proposed. Among these, the most studied, from both theoretical and experimental perspectives, are the semiconductor nanowires with high spin-orbit coupling in the presence of a magnetic field and in proximity to a $s$-wave superconductor \cite{lutchyn2010majorana, alicea2010majorana, sau2010generic}. The simplicity of the first theoretical models proposed, however, cannot account for all the physics underlying the electronic behavior of this system, and refined versions of the model are needed for comparison with experimental data. A possible phenomena is the appearance of subbands due to size quantization, similar to the discrete energies arising from quantum wells \cite{lutchyn2011search, stanescu2011majorana}. In the present work we show that for a nanowire with two subbands with the same helicity near the Fermi level, in the limit where spin-orbit coupling can be treated as a perturbation and the nanowire is weakly coupled to the $s$-wave superconductor, it is possible to map the system to a 2-band Kitaev chain. We also discuss the presence of chiral symmetries and its relevance to the computation of topological invariants as well as disorder effects caused by charged impurities.


\section{Subbands on finite size nanowires}

In order to describe a semiconductor nanowire with dimensions $L_z \ll L_y \ll L_x$ such that $L_z$ is negligible, we use the Hamiltonian \cite{lim2012magnetic}
\begin{equation}
\mathcal{H}_{SM} = \frac{1}{2m^*}(\Pi_x^2 + \Pi_y^2) - \mu + V(x,L_x;y,L_y),
\end{equation}
where $\Pi_i$ is the canonical momentum, $m^*$ is the effective mass, $\mu$ is the chemical potential and $V(x,L_x;y,L_y)$ is the confinement potential for the non-interacting electron gas. For a phenomenology similar to the Kitaev chain, we also need to introduce a Rashba spin-orbit coupling
\begin{equation}
\mathcal{H}_R = \lambda\left(\vec{\sigma}\wedge \vec{\Pi}\right)\cdot \hat{z},
\end{equation}
and a Zeeman field
\begin{equation}
\mathcal{H}_Z = \frac{1}{2}\vec{\sigma}\cdot\vec{B}.
\end{equation}

Finally, a proximity induced superconducting $s$-wave coupling can be added by taking the total unpaired Hamiltonian
\begin{equation}
\mathcal{H}_0 = \mathcal{H}_{SM} + \mathcal{H}_R + \mathcal{H}_Z
\end{equation}
and including a superconducting order parameter $\Delta$ according to the mean field Bogoliubov-de Gennes (BdG) prescription
\begin{equation}
\mathcal{H} = \left(
\begin{array}{cc}
\mathcal{H}_0 & \Delta \\
\Delta^* & - \mathcal{T}\mathcal{H}_0\mathcal{T}^{-1}
\end{array}\right),
\end{equation}
where $\mathcal{T} = i\sigma_y\mathcal{K}$ is the time-reversal operator and $\mathcal{K}$ denotes the complex conjugation operator.

Now, in order to treat the emergence of subbands in this system, it is necessary to  confine the electrons. Following the prescription given by Lutchyn \textit{et. al.} \cite{lutchyn2011search}, we take $L_x$ to infinity, separate the $x$ and $y$ coordinates and take the Zeeman field in the $x$ direction. Using $\tau_{\mu}$ and $\rho_{\mu}$ to denote electron-hole and subbands degrees of freedom, respectively, we have
\begin{align}
\mathcal{H} = \tau_z \otimes \left[ \left( \frac{p_x^2}{2m^*} - \mu \right) \sigma_0 \otimes \rho_0 - \lambda \sigma_y p_x \otimes \rho_0 \right. \nonumber \\ 
\left. + \frac{1}{2}\sigma_x B_x \otimes \rho_0 + \sigma_0 \otimes \frac{E_{sb}}{2}(\rho_0 - \rho_z) \right] \nonumber \\
- \tau_{\phi} \otimes \sigma_y \otimes \left( \rho_x |\Delta_{12}| + \rho_0 \Delta_+ + \rho_z \Delta_- \right) \nonumber \\
-E_{bm} \tau_0 \otimes \sigma_x \otimes \rho_y,
\label{red-ham}
\end{align}
where $\tau_{\phi} = \tau_x \sin \phi + \tau_y \cos \phi$, $E_{sb} = 3 \pi^2/2m^*L_y^2$ is the subband energy difference and $E_{bm} = 8 \lambda/3L_y$ is the band mixing energy. The superconducting order parameters
\begin{equation}
\Delta_{\pm} = \frac{|\Delta_1| \pm |\Delta_2|}{2}
\end{equation}
depend on the order parameter of the first and second subbands, $\Delta_1$ and $\Delta_2$, respectively. Also, $\Delta_{12}$ is the interband superconducting coupling. In the following, we will show that the Hamiltonian \eqref{red-ham} can be mapped to a 2-band Kitaev chain by treating the spin-orbit effects as a perturbation in the weak coupling limit for the induced superconducting order parameter.

\subsection{First order perturbation theory for the spin-orbit coupling}
To develop perturbation theory for the spin-orbit coupling, we take 
\begin{align}
\mathcal{H}_{0} = \tau_z \otimes \left[ \left( \frac{p_z^2}{2m^*} - \mu \right) \sigma_0 \otimes \rho_0 + \frac{1}{2}\sigma_z B_z \otimes \rho_0   \right. \nonumber \\
\left. + \sigma_0 \otimes \frac{E_{sb}}{2}(\rho_0 - \rho_z) \right]
\label{unpert}
\end{align}
as the unperturbed Hamiltonian, and
\begin{equation}\label{pert_ham}
\mathcal{H}_{pert} = \tau_z \lambda \sigma_y p_z \otimes \rho_0 - E_{bm} \tau_0 \otimes \sigma_z \otimes \rho_y
\end{equation}
as the perturbation. For later convenience, we performed a $\pi/2$ rotation around the $y$ axis to obtain equations \eqref{unpert} and \eqref{pert_ham} from \eqref{red-ham}. After a straightforward computation one concludes that the spectrum of the unperturbed system defined by \eqref{unpert} is
\begin{equation}
E^{(0)} = \pm \left[\left( \frac{p_z^2}{2m^*}-\mu\right) \pm \frac{1}{2}B + \left\lbrace \begin{array}{cc}
0 & \text{band 1} \\
E_{sb} & \text{band 2} \end{array}
\right. \right].
\label{eigenvalues}
\end{equation}
Next, assuming that the Zeeman field is strong enough for only one spin channel to be close to the Fermi level, say the spin-up component, we can calculate the first order correction to the set of eigenvectors corresponding only to the spin-up eigenvalues in \eqref{eigenvalues}.

\subsection{Effective superconducting pairing}
Having derived the first order corrections to the spin-up eigenvectors, we can compute the effective superconducting order parameters in the presence of a weak spin-orbit coupling:
\begin{align*}
\tilde{\Delta}_+ = \langle h, \uparrow, 1| \Delta_+ \tau_{\phi} \otimes \sigma_y \otimes \rho_0 | e, \uparrow, 1\rangle  =  ie^{i\phi} \frac{\lambda p_z}{B}\Delta_+,  \\
\tilde{\Delta}_- = \langle h, \uparrow, 2| \Delta_- \tau_{\phi} \otimes \sigma_y \otimes \rho_z | e, \uparrow, 2\rangle = ie^{i\phi}\frac{\lambda p_z}{B}\Delta_-, \\
\tilde{\Delta}_{12} = \langle h, \uparrow, 2| \Delta_{12} \tau_{\phi} \otimes \sigma_y \otimes \rho_x | e, \uparrow, 1\rangle = ie^{i\phi}\frac{\lambda p_z}{B}\Delta_{12}.
\end{align*}
Similar calculations can be performed to project all other terms of the Hamiltonian on the spin-up channel.

Finally, we project out the spin-down component by shifting the Fermi energy $\mu + B/2 \rightarrow \mu$ to obtain an effective spinless system, given by the following Hamiltonian:
\begin{align}\label{ham-eff}
\mathcal{H}_{eff} &= \tau_z \otimes  \left[ \left( \frac{p_z^2}{2m^*} - \mu \right)\rho_0 \nonumber + \frac{E_{sb}}{2}(\rho_0-\rho_z) \right] \\
 &+ \tau_0 \otimes E_{bm}\left[1-\left(\frac{\lambda p_z}{2B}\right)^2 \right]\rho_y \nonumber \\
&- \tau_{\phi} \otimes ie^{i\phi} \left( \rho_x |\Delta_{12}| + \rho_0 \Delta_+ + \rho_z \Delta_- \right)\frac{\lambda p_z}{B}.
\end{align}
The Hamiltonian \eqref{ham-eff} describes a spinless one-dimensional $p$-wave superconductor with two internal degrees of freedom corresponding to the two subbands of \eqref{red-ham}. Nonetheless, it is also possible to compare \eqref{ham-eff} with a spinful Kitaev chain. For the sake of the following discussion about topological invariants, it is convenient to define the quantities:
\begin{align}
\vec{d} &:= \frac{\lambda}{B}(-\Delta_-, \Delta_+, \Delta_{12}),\\
\epsilon(p_z) &:= \left( \frac{p_z^2}{2m^*} - \mu + \frac{E_{sb}}{2} \right),\\
\vec{V}(p_z) &:= \left(0, \mu(p_z), E_{sb}\right)
\end{align}
so that, after the similarity transformation
\begin{equation}
\psi = (c, c^{\dagger})^T \rightarrow \tilde{\psi} = (c, i\rho_2 c^{\dagger})^T,
\end{equation}
the effective Hamiltonian \eqref{ham-eff} becomes:
\begin{align}\label{repres}
\mathcal{H}_{eff} &= \epsilon(p_z) \tau_z \otimes \rho_0
+ \tau_0 \otimes \vec{V}(p_z) \cdot \vec{\rho} \nonumber \\
&- \tau_{\phi} \otimes e^{i\phi}\vec{\rho}\cdot \vec{d} \; p_z.
\end{align}

\section{Chiral symmetry and topological classification}
Next, to study the topological properties of the system and, therefore, properly define topological invariants to attest the presence of Majorana bound-states at both ends, it is necessary to account for all the discrete symmetries. Since charge conjugation ($\mathcal{C}$) is manifestly preserved by the BdG theory, the presence of a time-reversal (or at least a pseudo-time-reversal) symmetry $\mathcal{T}$ that squares to $+1\ (-1)$ implies the system is in BDI (DIII) class.

In fact, for a system with two internal degrees of freedom, there are two distinct topological classifications given by the chiral symmetries \cite{dumitrescu2015hidden}. Namely,
\begin{align} 
\label{chiraldiii} \mathcal{S}_{DIII} = \tau_{\phi+\pi/2} \otimes \rho_0,\\
\label{chiralbdi} \mathcal{S}_{BDI} = \tau_{\phi + \pi/2} \otimes \hat{d}\cdot \vec{\rho}.
\end{align}
The presence of chiral symmetry requires the Hamiltonian \eqref{repres} to commute with the chiral symmetry operator. Thus, we must have
\begin{equation}
\vec{V}(p_z) = 0
\end{equation}
for a DIII system, and
\begin{equation}
\vec{V}(p_z) \perp \vec{d}
\end{equation}
for a BDI system. In fact, the conditions for both topological classes are analogous to the results obtained by Dumitrescu \textit{et. al.} \cite{dumitrescu2015hidden} for a spinfull Kitaev chain where the term proportional to $\tau_0$ is the Zeeman (magnetic) field. We see that such term prohibits the system to be in DIII class, since it breaks a (pseudo) time reversal symmetry. For the BDI class, however, the condition is a little more flexible.

\begin{figure}[!ht]
\centering
\subfloat[\label{eigen-a}]{\includegraphics[width=.44\textwidth]{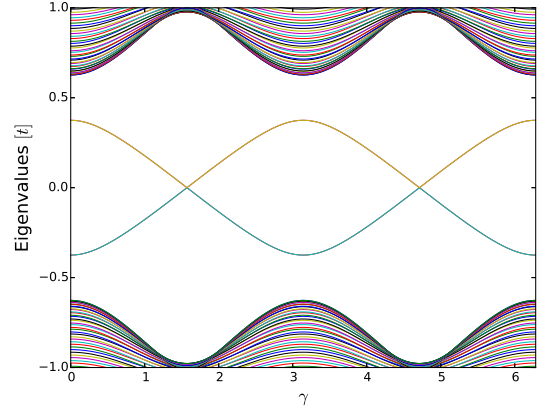}}\\
\subfloat[]{\includegraphics[width=.44 \textwidth]{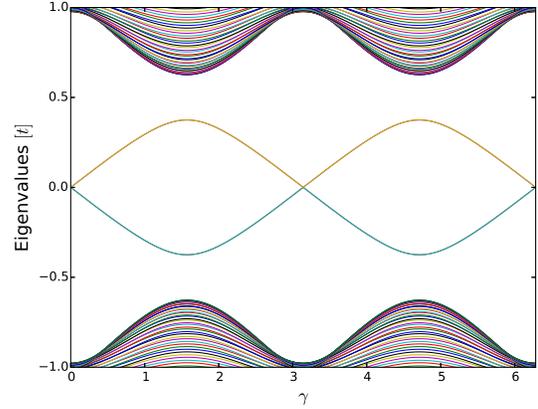}}
\caption{Eigenvalues (in units of $t$) for a system described by the Hamiltonian \eqref{numeric-1} with 100 lattice positions as a function of $\gamma$ with $\mu = 0$, $m = 0.5t$ and $d = 0.75t$. (a) $\rho_i = \rho_x, \rho_j = \rho_y$; (b) $\rho_i = \rho_y, \rho_j = \rho_z$. For $\rho_i = \rho_x$ and $\rho_j = \rho_z$ the spectrum is insensitive to changes in $\gamma$. Thus, it is evident that unless $\vec{d}$ lies on the $x-z$ plane, minigap states can emerge, indicating chiral symmetry breaking.}
\label{eigen-1}
\end{figure}

To corroborate the influence of the direction of the $\vec{d}$ vector on the presence of a non-trivial topological BDI phase, we performed independent numerical simulations with the Kwant package \cite{groth2014kwant}. For simplicity, we took the superconducting phase $\phi = 0$ and constrained $\vec{d}$ to lie on an arbitrary plane in band space with magnitude $d \equiv |\vec{d}| = 0.75t$, so that the Hamiltonian \eqref{repres} becomes:
\begin{align}\label{numeric-1}
\mathcal{H}_{eff} &= \epsilon(p_z) \tau_z \otimes \rho_0
+ m \tau_0 \otimes \rho_y\nonumber \\
&- \tau_{y} \otimes d\; \rho_{\gamma} p_z,
\end{align}
with $\rho_{\gamma} = \rho_i \sin \gamma + \rho_j \cos \gamma$, where $\rho_i$ and $\rho_j$ are arbitrary Pauli matrices in band space and $m$ is the momentum-independent term of $\mu_{12}(p_z)$ corresponding to the band mixing energy (the quadratic term in momentum was neglected). The results in Fig. \ref{eigen-1} clearly indicate that, unless $\vec{d}$ lies on the $x-z$ plane, \textit{i.e.}, $\vec{d}\perp \hat{y}$, minigap states proportional to the component of $\vec{d}$ along the $\hat{y}$ direction appear. This agrees with the condition for BDI chiral symmetry breaking.

\begin{figure}[!ht]
\centering
\includegraphics[scale=0.5]{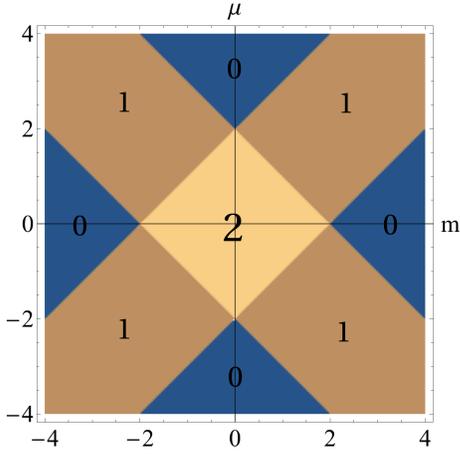}
\caption{Winding number calculated for the Hamiltonian  \eqref{numeric-1} varying $\mu$ and $m$, with $d = 0.75t$, $\rho_{\gamma} = \rho_x$. We see that three distinct topological phases are possible, with winding number 0, 1, and 2.}
\label{winding}
\end{figure}

Finally, we can introduce a topological invariant for the BDI class, the winding number, $w \in \pi_1(S^1)=\mathbb{Z}$ \cite{sato2011topology,tewari2012topological}, defined by:
\begin{equation}
w = \left| \oint_{BZ} \frac{dk}{4\pi i} \text{tr}[\mathcal{S}_{BDI}H^{-1}_k\partial_k H_k] \right|,
\end{equation}
to count the number of Majorana bound states hosted at the ends of the nanowire. In Fiq. \ref{winding} we show the resulting topological phase diagram as a function of the chemical potential $\mu$ and the band mixing $m$ for a case in which the Hamiltonian \eqref{numeric-1} preserves BDI chiral symmetry. A winding number equal to 2 indicates that two pairs of Majorana excitations are present, one pair for each subband.

\section{Disorder effects}

Finally, we investigate the robustness of the system against disorder generated by charged impurities. This can be done by changing the chemical potential in \eqref{numeric-1} as
\begin{equation} \label{disorder-mu}
\mu \rightarrow \mu(z) = \mu + U(z),
\end{equation}
where the local disorder potential is given by \cite{wurm2012symmetries}:
\begin{equation}
U(z) = \sum_i U_i \exp \left(-\frac{(z-z_i)^2}{2\xi^2}\right),
\end{equation}
with a random distribution of charged impurities at the positions $\{z_i\}$, a characteristic length associated with the disorder range $\xi$, and a function $U_i$ with random values between $-U_0/2$ and $U_0/2$.

\begin{figure}
\subfloat[\label{eigen-a}]{\includegraphics[scale=.22]{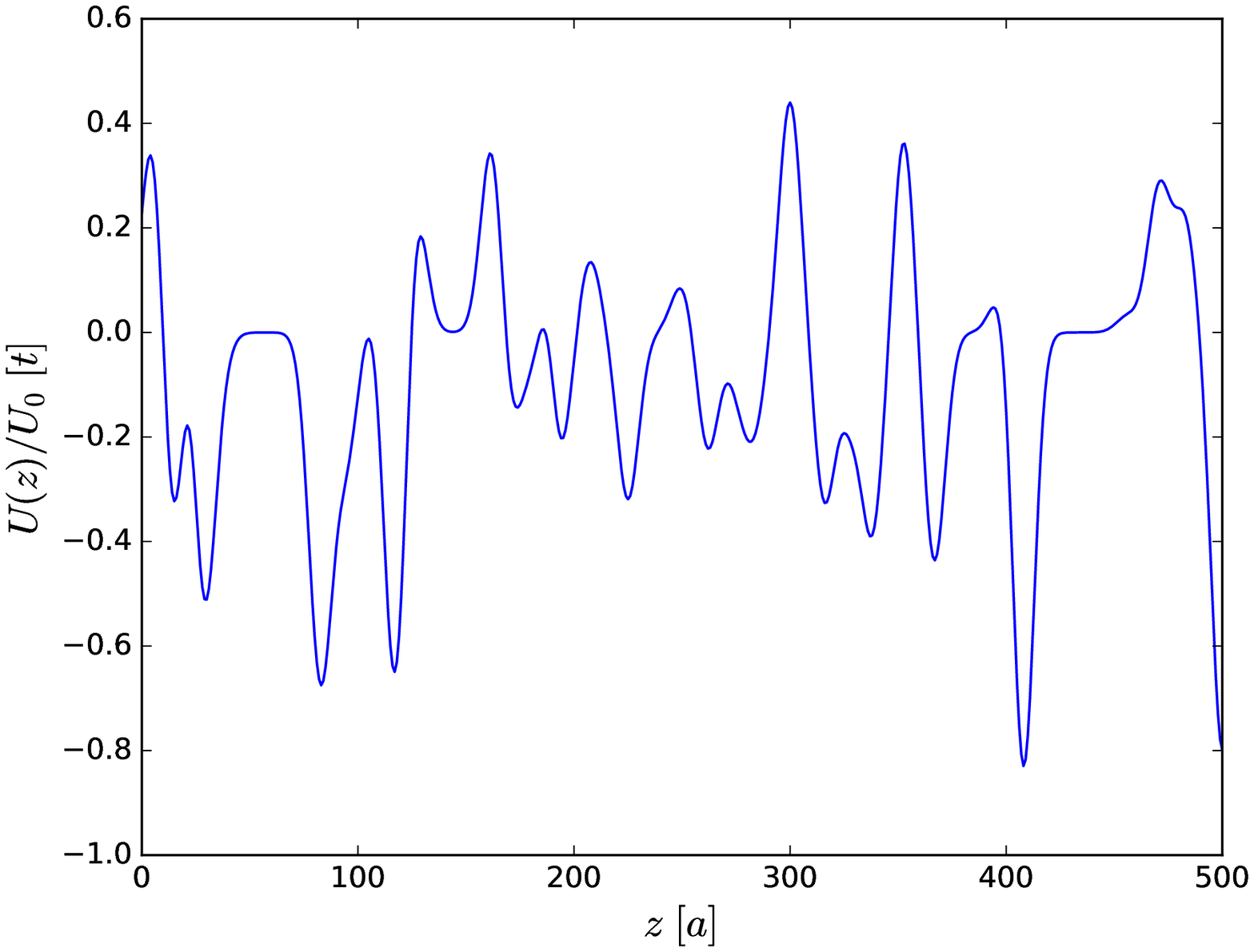}}
\subfloat[]{\includegraphics[scale=.22]{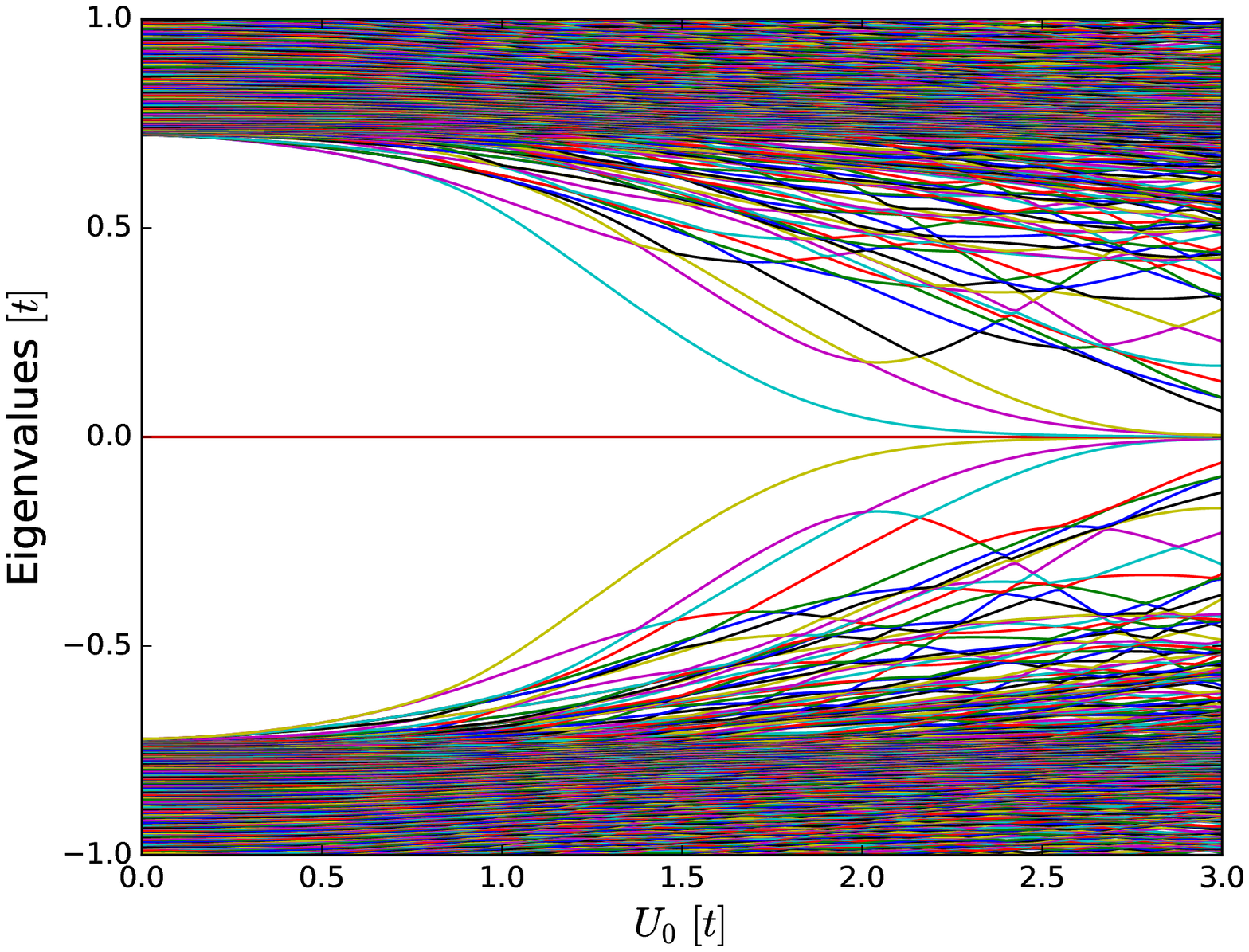}}\\
\subfloat[\label{eigen-a}]{\includegraphics[scale=.22]{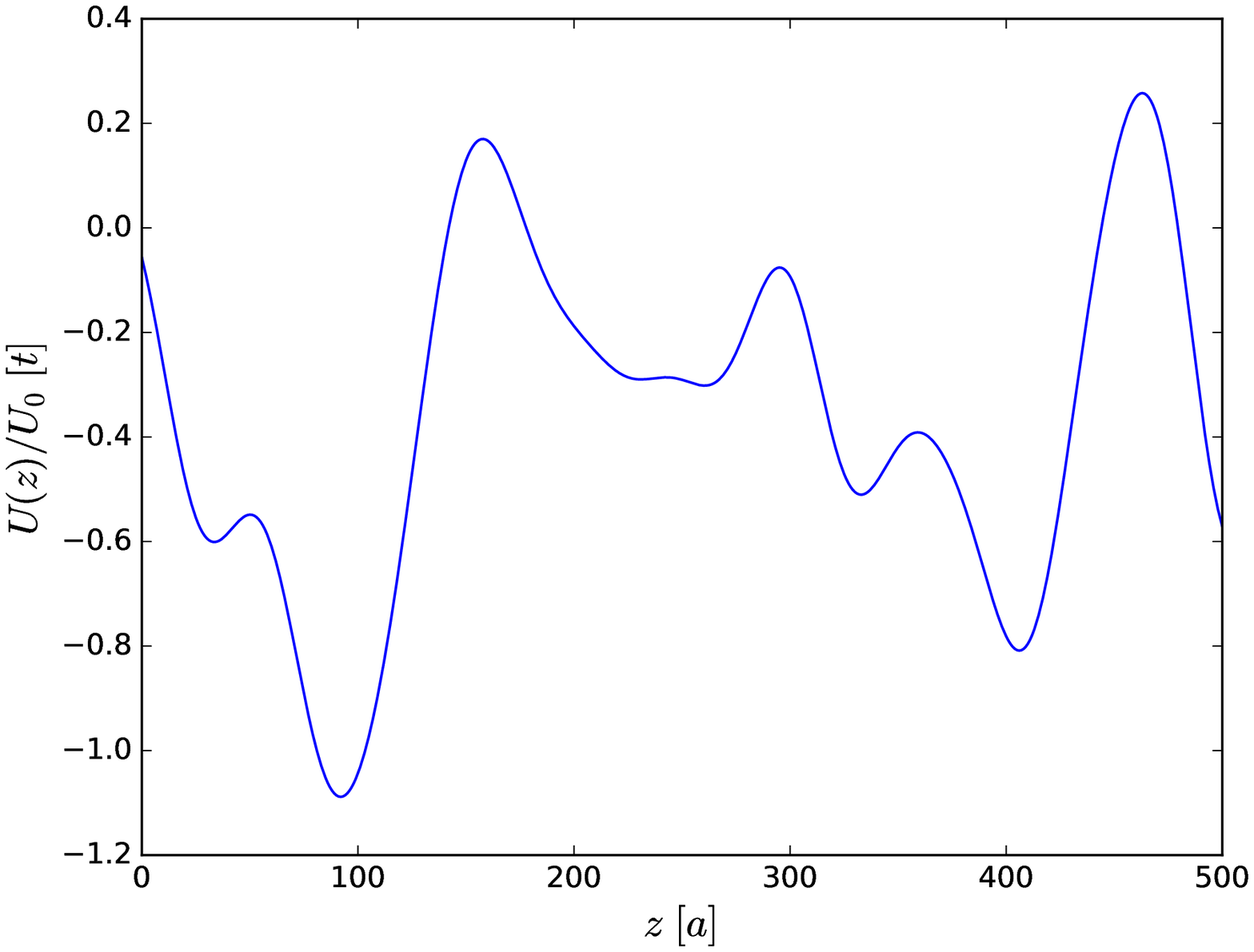}}
\subfloat[]{\includegraphics[scale=.22]{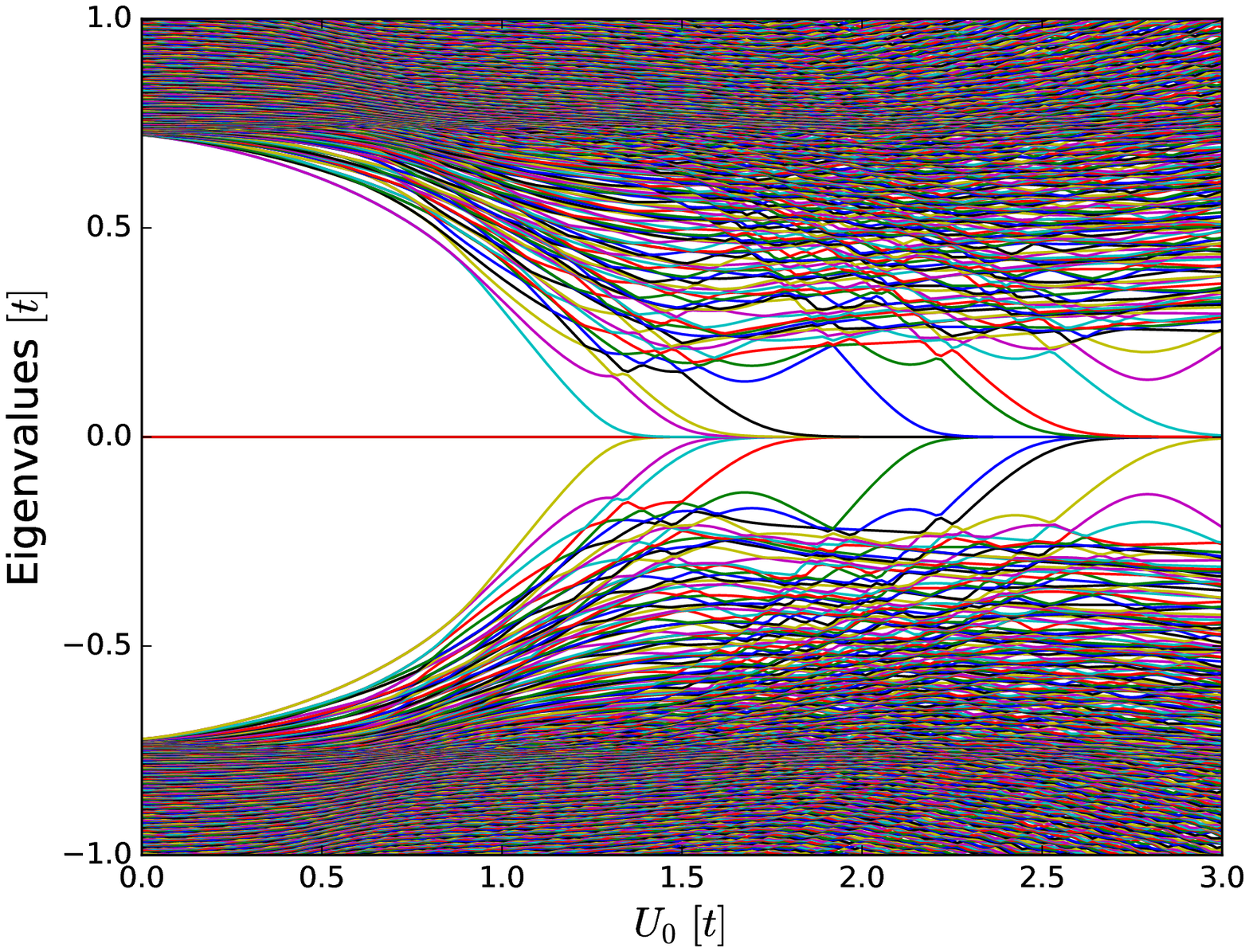}}\\
\subfloat[\label{eigen-a}]{\includegraphics[scale=.22]{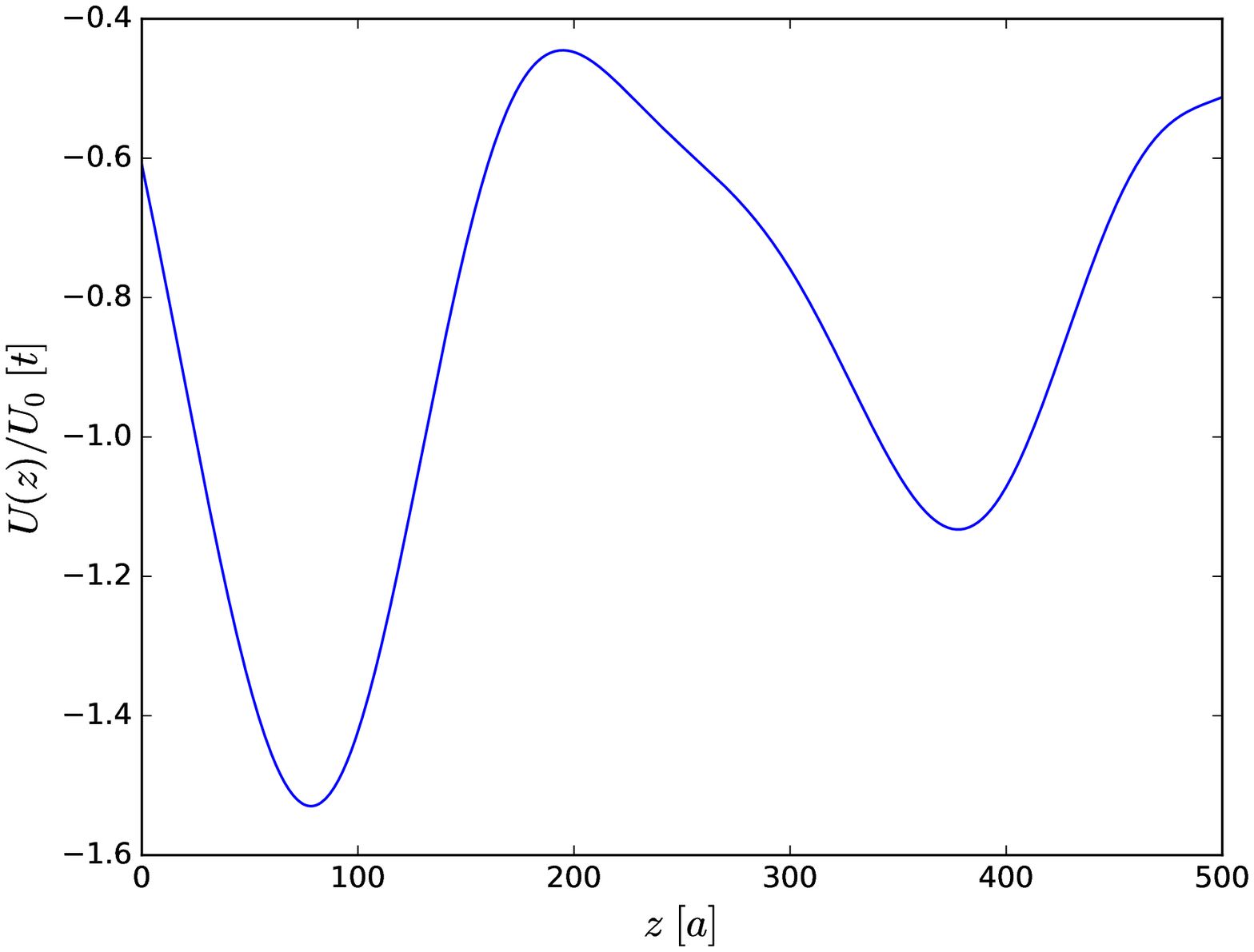}}
\subfloat[]{\includegraphics[scale=.22]{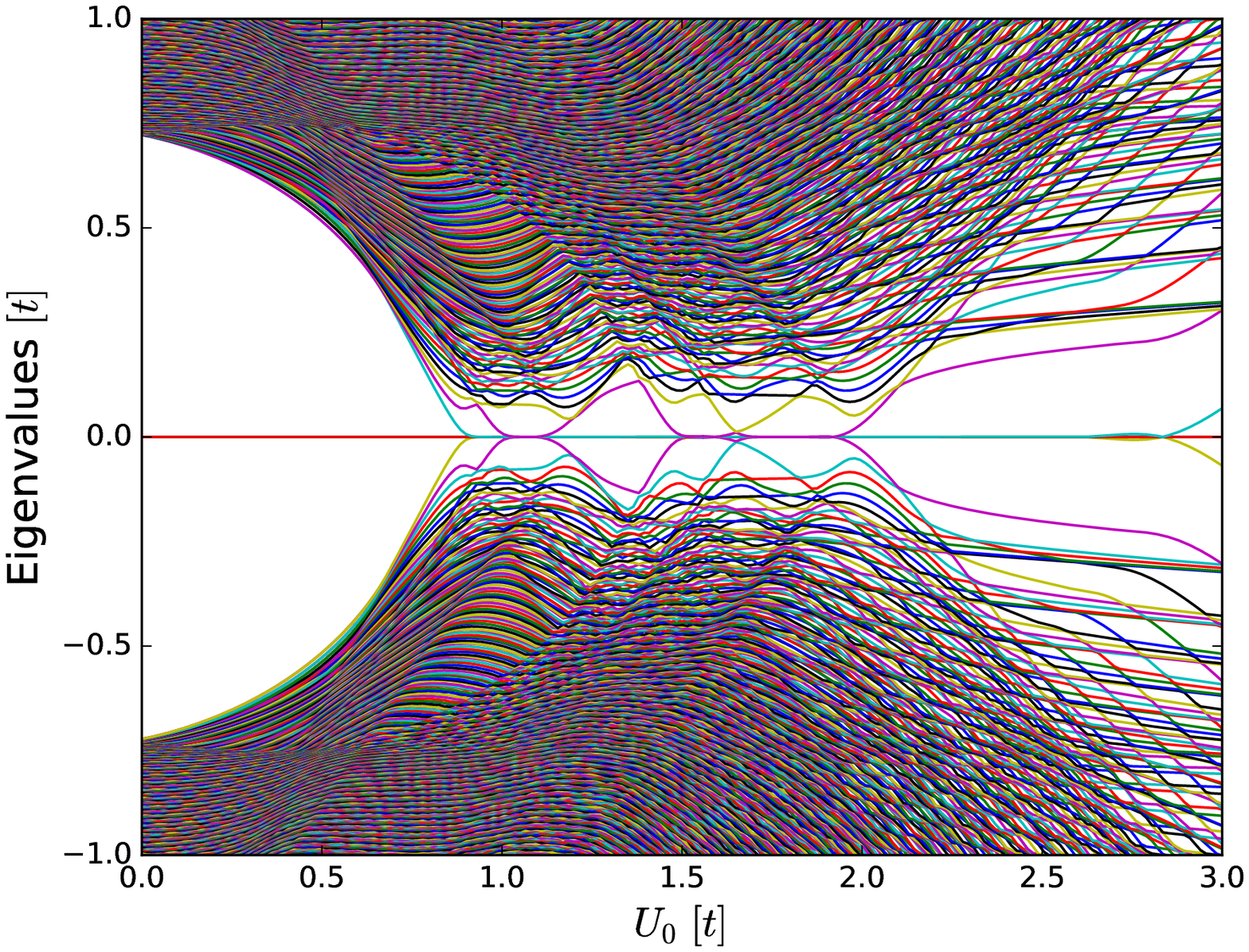}}
\caption{Disorder profiles (a), (c) and (e) and the corresponding evolution of the spectrum with increasing disorder amplitude (b), (d) and (f), respectively. Results obtained for the Hamiltonian in Eq. \ref{numeric-1} in a chain with $\mu = 0$, $d = 0.75t$, $m = 0.5t$ and $\gamma=0$ in a lattice with 500 positions and 50 charged impurities for three different disorder ranges: (a,b) $\xi = 5a$, (c,d) $\xi = 20a$, and (e,f) $\xi = 50a$.}
\label{disorder}
\end{figure}

In Fig. \ref{disorder} we show the effect of increasing the disorder amplitude $U_0$ in the eigenenergies spectrum, as well as the influence of the characteristic length of the interaction on the Hamiltonian of \eqref{numeric-1} with the modified chemical potential \eqref{disorder-mu}. Although, zero modes persist for low disorder amplitudes due to the topological nature of the system, increasing the disorder amplitude eventually leads to the collapse of the bulk gap. Clearly, the longer the range of interactions are, the more effective the potential is to close the bulk gap. A similar behavior is observed for finite size nanowires \cite{lutchyn2011search}. This can be explained as follows. For short-range disorder, a pair of Majorana bound states emerge at the charged impurities positions and hybridize forming subgap states. On the other hand, increasing the disorder range leads to larger separations between the emerging Majorana bound states and an overlap between the wavefunctions of the Majorana states at the ends and at disorder positions becomes increasingly possible.

\section{Conclusion}

We have shown that there is a mapping between finite size semiconductor nanowires with high spin orbit coupling and induced $s$-wave superconductivity to a 2-band Kitaev chain, \textit{i.e.}, a spinless 2-band $p$-wave superconductor. This is true provided that there are two subbands near the Fermi level and the Zeeman splitting is sufficiently strong to guarantee just one helicity in the limit of weak superconducting coupling and if the spin-orbit effects are small enough to be treated as perturbations. The results are equivalent to spinful $p$-wave superconductors, as there is a relation between the effects of the interband coupling and the Zeeman field on the breaking of chiral symmetry. Furthermore, the topological phase diagram corroborates the results on the literature regarding the possibility of existing multiple Majorana excitations at both ends of finite size nanowires when BDI chiral symmetry is preserved. We remark that this can be a suitable route for detecting more intense Majorana zero-bias conductance peak in nanowires.

Finally, we have shown that the collapse of the bulk gap as a result of disorder generated by charged impurities does not depend only on the disorder amplitude, but it is also related to the range of the disorder. In other words, increasing the characteristic length of the disorder generated by charge fluctuations results in gap closings for lower values of the disorder amplitude.


%





\ifCLASSOPTIONcaptionsoff
  \newpage
\fi

\newpage


\bibliographystyle{IEEEtran}
\bibliography{nanowire}
\end{document}